\documentclass[runningheads]{llncs}
\usepackage[T1]{fontenc}
\usepackage{float}
\usepackage{graphicx}
\usepackage{url}
\usepackage{xcolor}
\usepackage{multirow}
\usepackage{adjustbox}

\usepackage{graphicx}
\usepackage{colortbl}

\begin{document}

\title{Can we predict QPP? An approach based on multivariate outliers}
%
%

\author{Adrian-Gabriel Chifu\inst{1}\orcidID{0000-0003-4680-5528}
\and
Sébastien Déjean\inst{2}\orcidID{0000-0001-9610-5306}
\and
Moncef Garouani\inst{4}\orcidID{0000-0003-2528-441X}
\and
Josiane Mothe\inst{3,5}\orcidID{0000-0001-9273-2193}
\and
Diégo Ortiz\inst{5}\orcidID{0009-0002-7272-5644}
\and
Md Zia Ullah\inst{6}\orcidID{0000-0002-4022-7344}
}

\authorrunning{A.-G. Chifu et al.}
%
\institute{
Aix Marseille Université, Université de Toulon, CNRS, LIS, Marseille, France
\email{adrian.chifu@univ-amu.fr}\\
\and
IMT, UMR5219 CNRS, UPS, Univ. de Toulouse, Toulouse, France
\email{sebastien.dejean@math.univ-toulouse.fr}\\
\and
IRIT, UMR5505 CNRS, Université de Toulouse, INSPE, UT2J, Toulouse, France
\email{josiane.mothe@irit.fr}\\
\and
IRIT, UMR 5505 CNRS, Université Toulouse Capitole, UT1, Toulouse, France
\email{moncef.garouani@irit.fr}\\
\and
IRIT, UMR 5505, Toulouse, France
\email{diego.ortiz@irit.fr}\\
\and
Edinburgh Napier University, Edinburgh, UK
\email{m.ullah@napier.ac.uk}\\
}

\maketitle              
\begin{abstract}
Query performance prediction (QPP) aims to forecast the effectiveness of a search engine across a range of queries and documents. While state-of-the-art predictors offer a certain level of precision, their accuracy is not flawless. Prior research has recognized the challenges inherent in QPP but often lacks a thorough qualitative analysis. In this paper, we delve into QPP by examining the factors that influence the predictability of query performance accuracy. We propose the working hypothesis that while some queries are readily predictable, others present significant challenges.
By focusing on outliers, we aim to identify the queries that are particularly challenging to predict. To this end, we employ multivariate outlier detection method. Our results demonstrate the effectiveness of this approach  in identifying queries on which QPP do not perform well, yielding less reliable predictions. Moreover, we provide evidence that excluding these hard-to-predict queries from the analysis significantly enhances the overall accuracy of QPP.

 \and Multivariate outlier detection
\end{abstract}
\section{Introduction}

A search engine aims to process and answer any user query by retrieving relevant documents. However, the performance of a particular search engine can vary substantially depending on the specific queries it encounters~\cite{harman2009overview,deveaud2018learningLong}.

Query performance prediction (QPP) addresses the crucial task of predicting how effective a system will be on a given query~\cite{zhou2007query,carmel2010estimating,shtok2012predicting,chifu2018query}. This problem is of paramount importance for two primary reasons. Firstly, when a query is expected to be difficult, it may be necessary to adopt a specialized, albeit potentially costly, approach. Conversely, when a query is predicted to be easy, a simpler and more cost-effective method can be employed~\cite{CronenTownsend2002,amati2004query,dejean2019studying}.

QPP accuracy is typically assessed by measuring the correlation between the predicted and the actual performance values~\cite{shtok2012predicting,raiber2014query,faggioli2022smare,datta2022pointwise}. This approach is sound as long as the underlying assumptions and conditions for applying correlation measurements are respected~\cite{anscombe1973graphs}. 

Systems vary in their approach to processing a given query. This variability encompasses processes such as automatic query reformulation, the choice of the weighting/matching function (BM25~\cite{robertson2009probabilistic}, a Language model~\cite{zhai2004study}, or an LLM-based model~\cite{devlin2018bert,liu2019roberta,clark2020electra,zhu2023large}, for example), and the application of document re-ranking models~\cite{liu2009learning}. Consequently, different systems will perform differently on the same queries~\cite{dejean2019studying}. QPP is thus considered with regard to the system it predicts the performance for. This implies that a QPP predictor should adapt its behavior to suit a particular system. This could be a reason why post-retrieval QPPs, which use the results of the search, tend to be more accurate than pre-retrieval ones, which rely solely on the query and the set of documents~\cite{hauff2008survey,JAFARZADEH2022102746}.
 
\begin{figure}[!ht]
\vspace{-5pt}\begin{center}
    \includegraphics[scale=0.9]{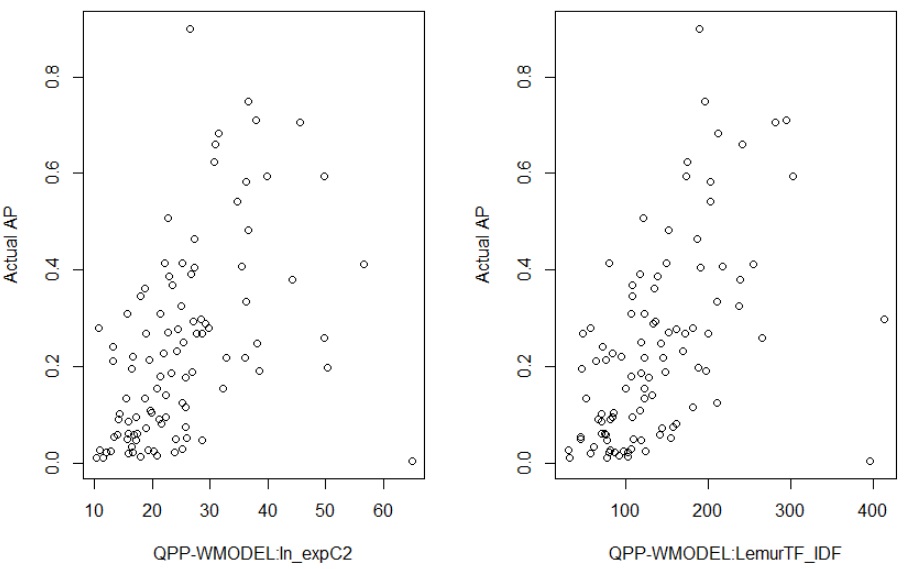}
\end{center}
\vspace{-5pt}
\caption{\textbf{QPPs behave differently on the same system and set of queries (TREC78).} Predicted AP (X-axis) and actual AP (Y-axis) obtained with LGD weighting function and their Pearson correlation. Left side {In\_expC2} ($\rho$=0.484) - Right side {LemurTF\_IDF} QPP ($\rho$=0.552). } \label{Fig:Ref_MAP_Letor_TREC78}
\vspace{-5pt}
\end{figure}

We found for example that \textit{LemurTF\_IDF}, 
a post-retrieval QPP corresponding 
to a Letor feature~\cite{cao2007learning,chifu2018query} has a higher correlation with the actual  Average precision (AP) obtained with the LGD weighting function~\cite{clinchant2010information} than with the JS weighting function~\cite{amati2006frequentist}  (resp. 0.522 and 0.504 Pearson correlation). On the other hand, two predictors will behave differently on a particular system (Fig.~\ref{Fig:Ref_MAP_Letor_TREC78}).   

In Figure~\ref{Fig:Ref_MAP_Letor_TREC78}, we can see that the correlation is relatively weak, which is accurately reflected by the 0.484 (resp. 0.552) Pearson correlation values. Current QPP models lack accuracy. Single features, even post-retrieval ones could not demonstrate high correlation with actual performance~\cite{hauff2008survey,carmel2010estimating,faggioli2023query}; even the combination of predictors, which is out of the scope of this paper, has not been very successful~\cite{grivolla2005automatic,zhou2007query,raiber2014query,chifu2018query,mizzaro2018query,ROY20191026}.

In this paper, we aim to analyze  performance prediction deeper. Past studies showed that predictors do not work well when considering all queries. We hypothesize that prediction accuracy can significantly vary among queries, and our objective is to identify the queries for which the predictor fails to estimate the effectiveness.
In other words, we want to predict the predictability of performance and address the following research question: 
Is it possible to predict the queries for which a QPP can provide an accurate effectiveness estimation for a given system?
To put it differently, can we concentrate on those queries that are more likely to yield accurate predictions and perhaps automatically disregard queries that are more challenging to predict?

Some queries can be considered outliers (\textit{abnormally} easy or difficult); similarly, predictions may have abnormal values. We hypothesize that the queries with abnormal prediction values are difficult to predict or get unreliable predictions. To tackle this problem, we consider multivariate outlier detection as a means to identify these hard-to-predict queries. Since a given QPP may behave differently to estimate the effectiveness of a system, the idea is to consider multiple QPPs in the query identification phase. 

We consider several effectiveness measures and several benchmark collections. We show that our hypothesis could pave the way for a new research direction on QPP on accuracy predictability. 

\section{Identification of difficult to predict queries by TRC}
Here, we hypothesize that some predictions are outliers because the queries are difficult to predict. We aim to identify those queries for which we anticipate the QPP will not be accurate. 

Johnson defines an outlier as an observation in a data set that appears to be inconsistent with the remainder of that set of data~\cite{johnson2002applied}. Univariate methods consider each variable independently, so only observations that appear odd for that variable are detected, while in the case of multivariate outlier detection, the interactions among
different variables are compared. Multivariate outliers are a combination of unusual scores on several variables~\cite{pena2001multivariate}, and the idea is to detect the observations that are located relatively far from the center of the data distribution~\cite{ben2005outlier}. 
To identify outliers, we thus consider here multivariate outliers and identify the queries for which predictions are abnormal for different QPPs. 

Mahalanobis distance is a common criterion for multivariate outlier detection. Applied to QPP, the Mahalanobis distance for a given query $q_i$, from a set of $n$ queries $Q = \{q_1, q_2, \dots, q_n\}$, can be defined as follows:

\vspace{-5pt}\begin{equation}
D_{M}\left(q_{i}, Q\right) = \sqrt{(q_{i} - \overline{q})^{T} V_{n}^{-1} (q_{i} - \overline{q})} ,
\end{equation}
\noindent where the query vector $q_i$ is composed of $m$ ($n >> m$) predictor values $q_i = \left(p_1^i, p_2^i, \dots, p_m^i\right)$, $\overline{q}$ is the mean vector of the queries, and $V_{n}^{-1}$ is the inverse of the covariance matrix of the queries. The superscript $T$ denotes the transpose of the vector.
We used the Transformed Rank Correlations (TRC) for multivariate outlier detection  function implemented in R in this paper~\cite{beguin2004multivariate}\footnote{\url{https://cran.r-project.org/web/packages/modi/modi.pdf}}. Instead of working with the raw data, which might be skewed or have a non-normal distribution, the TRC method works with the ranks of the data. This involves sorting the data from smallest to largest and using their position in this sorted order as their value. It then calculates the Spearman rank correlation or a similar non-parametric correlation measure between all pairs of variables. This helps in understanding the underlying structure and relationships in the data, which is crucial for identifying outliers in a multivariate context. The TRC method calculates a distance measure for each observation in the multivariate space. This distance is typically based on rank correlations and reflects how far away each observation is from the central trend of the data.  It then determines a threshold for these distances (default value is set to the 0.95 quantile of F-distribution). Observations with a distance greater than this threshold are flagged as outliers. 

\section{Data}\label{sec:data}

In this study, we use  TREC adhoc collections: TREC78 with 528K documents and 100 queries (351 -- 450) and WT10G which consists of 1.692M documents and 100 queries  (451 -- 550).
We use Average Precision (AP@1000)~\cite{sakai2007reliability} and normalized Discounted Cumulative Gain (nDCG@20)~\cite{jarvelin2017ir}  as measures to evaluate the search engine performance, which are common in adhoc retrieval. Like previous studies on QPP~\cite{grivolla2005automatic,zhou2007query,carmel2010estimating,shtok2012predicting,raiber2014query,chifu2018query}, we use Pearson correlation as a measure to evaluate the accuracy of the QPP, in addition to plots as recommended in statistics. Pearson correlation measures the linear correlation between the predicted effectiveness of queries and their actual effectiveness. It indicates how well the QPP method predictions align with the actual query performances. Although it may not be a perfect measure~\cite{Mothe202329}, we kept it in this study. Alternative measures will be used in future work. We will consider other measures such as Mean Squared Error (MSE), Mean Absolute Error (MAE) or its normalized variant. The former measures the average of the squares of the errors between predicted and actual query effectiveness. The latter measures the average of the absolute differences between predicted and actual values. It is less sensitive to outliers compared to MSE. These two measures could help in understanding the magnitude of the prediction errors.

We consider a series of systems that treat the same queries over the same set of documents. 
In this study, we construct systems using Terrier~\cite{ounis2005terrier}. These systems differ  on several factors: the scoring function employed and the variant of the query reformulation module utilized, if any.
We report the results on the best system according to the considered collection and performance measures, the best in terms of the average effectiveness over the set of queries. We also used a reference system based on LM Dirichlet~\cite{zhai2004study}.

%
In this study, we consider the first four Letor features which have been shown as among the most accurate for single feature QPP~\cite{cao2007learning,chifu2018query}. Letor features have been initially used for retrieved document re-ranking~\cite{cao2007learning}. Letor features are associated with each (query, retrieved document) pair~\footnote{using Terrier (see \url{http://terrier.org/docs/v5.1/learning.html})}. To obtain a single value for each query for a given letor feature, the values are aggregated over the documents. We used maximum as the aggregation function which has been shown as the most accurate for QPP~\cite{chifu2018query}. The four features we kept are LemurTF\_IDF, In\_expC2, InB2, and InL2, aggregated using the maximum function. 
We also consider four state-of-the-art QPP: Normalized query commitment (NQC)~\cite{shtok2012predicting}, Unnormalized query commitment (UCQ)~\cite{shtok2012predicting}, QF (query feedback)~\cite{zhou2007query} and WIG~\cite{zhou2007query} .

These features will be used individually to predict system performance. Although methods that combine features have been shown more accurate, here the objective is to conduct a first study on how to detect queries that are difficult to predict. We though opted for simple prediction models (single feature); we will analyze the proposed method for more complex models. Complex models that involve several features need to be trained. Here since we are not training a model, we do not need to split the data into train and test. 

\section{Results}\label{sec:results}

\begin{figure}[!ht]
\begin{adjustbox}{width=\textwidth}
\includegraphics{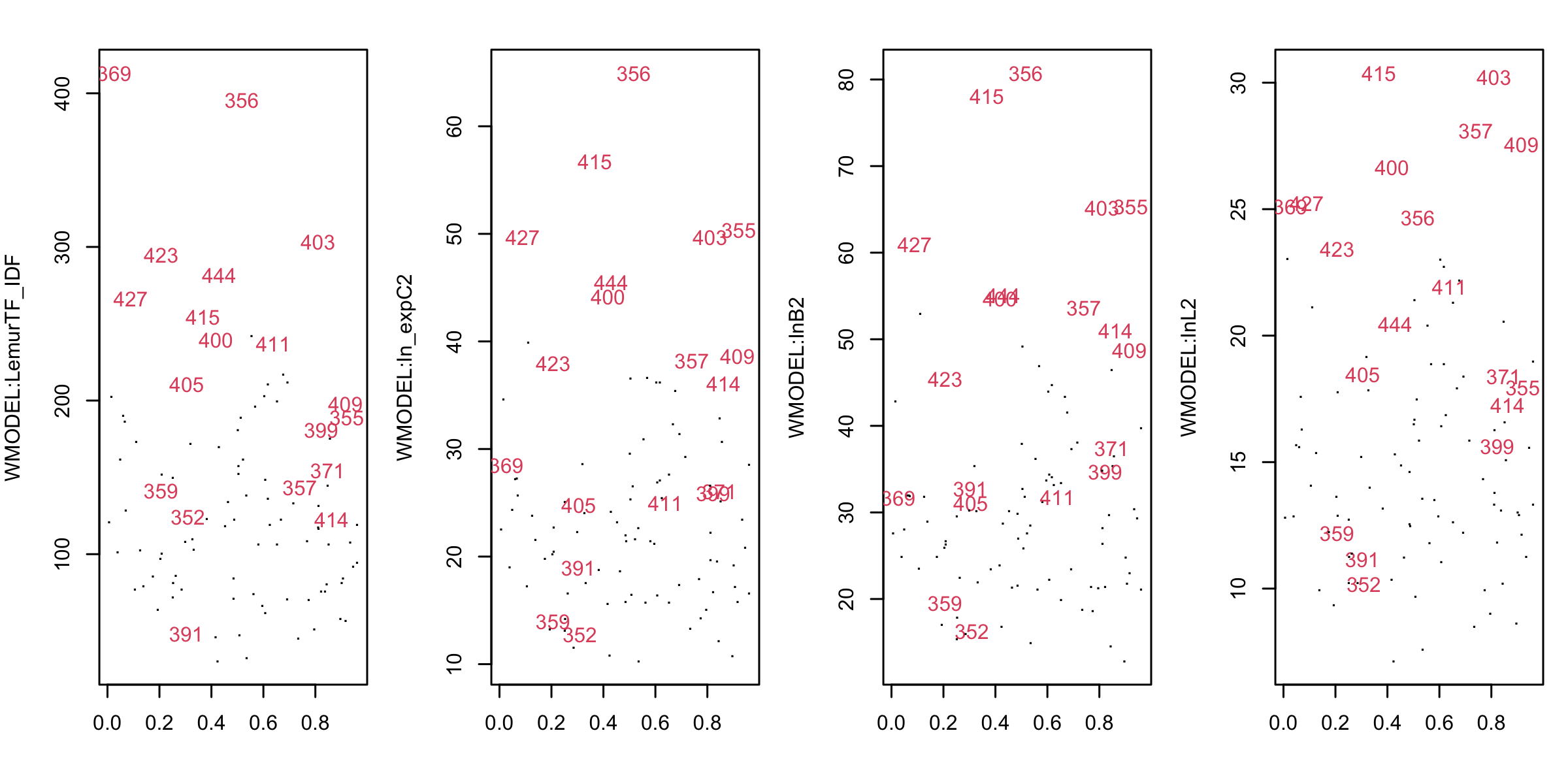}\end{adjustbox}
\vspace{-5pt}
\caption{\textbf{Outlier queries are different according to the QPP.} Queries  detected as outliers by our method are  in red. The four plots correspond to the four QPPs. The X-axis is not meaningful. 
Y-axis corresponds to the predicted values by the considered Letor QPP. QQP values are calculated for the TREC78 collection.
\vspace{-5pt}\label{Fig:Multi outliers}}
\end{figure}

\begin{figure}[!ht]
\begin{adjustbox}{width=\textwidth}
\includegraphics{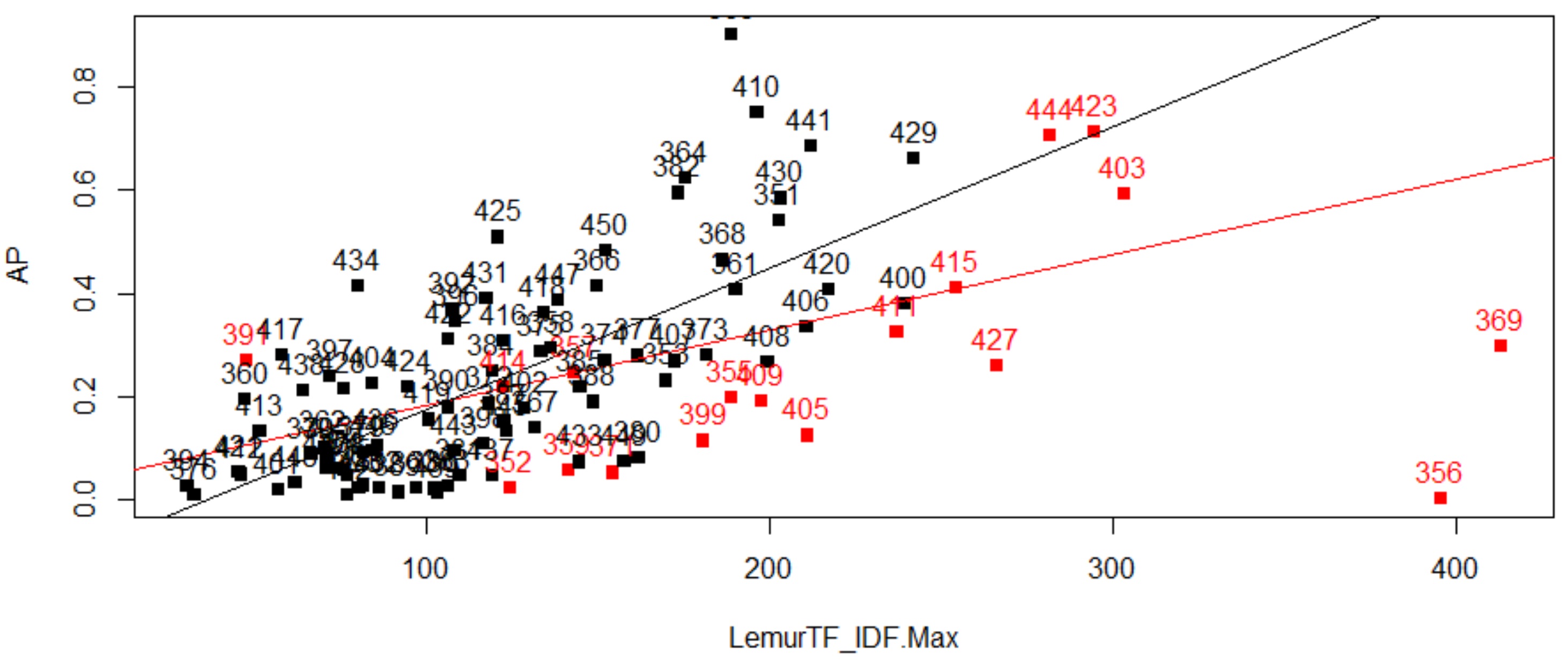}\end{adjustbox}
\vspace{-5pt}
\caption{\textbf{Outlier queries are correctly identified.} Queries that have been detected as outliers by our method are displayed in red. X-axis represents the QPP values and Y-axis is the actual AP. Here the values are calculated for LemurTF\_IDF on the TREC78 collection and a model based on the LGD weighting function. 18 out of 100 queries were identified as outliers. The red line is the regression when all the topics are considered while the black one is after the outliers were removed.}
 \label{Fig:TREC78_AP_21_Letor}
\end{figure}

Outlier queries are different according to the considered QPP, which shows the importance of using multivariate outlier detection (Fig.~\ref{Fig:Multi outliers}). For example, query 403 is a clear outlier for InL2 QPP (top right of the right-side sub-figure) but not for the other QPPs.
This implies that, if we used a univariate method such as LemurTF\_IDF in isolation, we would not have identified this query as one that is difficult to predict, even though it truly is (Fig.~\ref{Fig:TREC78_AP_21_Letor}).  Similarly, consider Query 369 (shown at the top left in the left-side sub-figure). It clearly stands out as an outlier when evaluated using the LemurTF\_IDF QPP, but does not exhibit the same outlier behavior when assessed by the other three QPPs.
These two queries (360 and 403) are indeed not well predicted (Fig.~\ref{Fig:TREC78_AP_21_Letor}). In addition, we can see that query 356 has also been accurately identified as an outlier, as well as some other queries (Fig.~\ref{Fig:TREC78_AP_21_Letor}). If these queries were not considered when calculating the correlation, the correlation would be higher. 

We can observe that the correlation when considering the outliers only is very weak (Tab.~\ref{Tab:Results}, ``LemurTF\_IDF - Outliers only" row, left-part). This result was expected since we want to remove these difficult to predict queries. On the contrary, when the outlier queries are removed, the correlation between the QPPs and the actual effectiveness measures is much higher (Tab.~\ref{Tab:Results} ``No Outliers" rows). Note that if we use univariate outlier detection using LemurTF\_IDF predictor only, the correlation decreases (0.658 compared to 0.700 here) (Tab.~\ref{Tab:Results}  ``Univariate" rows). 
Results are consistent across the QPPs for TREC78 collection. This 
consistency remains valid for the best predictor LemurTF\_IDF also for WT10G. We also evaluated the results considering other reference systems (NN-based model will be included in future work) and the results were also consistent (e.g., using In\_expB2 weighting function).
Results are also consistent when considering the other 4 QPP from the literature. Here we consider one single reference system (LM-based), WT10G and TREC Robust collections (Tab.~\ref{Tab:Results}, right side part).

\vspace{-5pt}
\begin{table} [!ht]
\caption{\textbf{Pearson correlation is consistently better when queries our method detects as difficult to predict are removed}. Correlation between actual effectiveness (either AP or ndcg) on  TREC78 and WT10G. We report the number of outlier queries detected, the correlation when outliers are removed using univariate and multivariate methods and when all the queries are considered for the 4 Letor features and 4 other features. The reference system rows indicate the effectiveness on average for the considered measure and set of queries; this is the system that obtained the highest effectiveness measure over a set of system configurations we tried. A unilateral test of significance for the difference between the correlations with ``No Outliers" and ``All" were performed using the cocor R package \cite{Diedenhofen2015}. A * after a correlation indicates that the p-value of the test is lower than 0.05. } \resizebox{\textwidth}{!}{%
\begin{tabular}{
>{\columncolor[HTML]{FFFFFF}}l 
>{\columncolor[HTML]{FFFFFF}}c 
>{\columncolor[HTML]{FFFFFF}}c 
>{\columncolor[HTML]{FFFFFF}}c 
>{\columncolor[HTML]{FFFFFF}}c}
\hline
\multicolumn{1}{c}{\cellcolor[HTML]{FFFFFF}Collection} & WT10G          & WT10G          & TREC78         & TREC78           \\ 
\multicolumn{1}{c}{\cellcolor[HTML]{FFFFFF}Measure}    & NDCG           & AP             & NDCG           & AP               \\
\multicolumn{1}{c}{\cellcolor[HTML]{FFFFFF}Reference system}& 0.444          & 0.236          & 0.524          & 0.238            \\
\multicolumn{1}{c}{\cellcolor[HTML]{FFFFFF}Outliers}   & 16             & 16             & 19             & 18               \\ \hline
LemurTF\_IDF - Univariate                              & 0.337          & 0.342          & 0.544          & 0.658            \\
LemurTF\_IDF - Outliers only                           & 0.206          & 0.292          & 0.095          & 0.350            \\
LemurTF\_IDF - No Outliers                             & \textbf{0.438} & \textbf{0.468} & \textbf{0.601}* & \textbf{0.700}*   \\
LemurTF\_IDF - All                                     & 0.365          & 0.393          & 0.381          & 0.522            \\ \hline
In\_expC2 - Univariate                                 & 0.423          & 0.368          & 0.607          & 0.631            \\
In\_expC2 - No Outliers                                & 0.391          & 0.350          & \textbf{0.607}* & \textbf{0.635}   \\
In\_expC2 - All                                        & 0.425          & 0.371          & 0.418          & 0.484            \\ \hline
InB2 - Univariate                                      & 0.329          & 0.286          & 0.542          & 0.536            \\
InB2 - No Outliers                                     & 0.286          & 0.214          & 0.530          & \textbf{0.543}   \\
InB2 - All                                             & 0.336          & 0.274          & 0.372          & 0.416            \\ \hline
InL2  - Univariate                                     & 0.264          & 0.341          & 0.380          & 0.426            \\
InL2 - No Outliers                                     & 0.258          & 0.347          & \textbf{0.458} & \textbf{0.491}   \\
InL2 - All                                             & 0.340          & \textbf{0.353} & 0.398          & 0.446            \\ 
\end{tabular}
\begin{tabular}{lcccc}
\hline
\rowcolor[HTML]{FFFFFF} 
\multicolumn{1}{c}{\cellcolor[HTML]{FFFFFF}Collection} & WT10G                & WT10G                & TREC78               & TREC78               \\ 
\rowcolor[HTML]{FFFFFF} 
\multicolumn{1}{c}{\cellcolor[HTML]{FFFFFF}Measure}    & NDCG                 & AP                   & NDCG                 & AP                   \\
\rowcolor[HTML]{FFFFFF} 
\multicolumn{1}{c}{\cellcolor[HTML]{FFFFFF}Ref system} & 0.4528               & 0.187                & 0.5235               & 0.2538               \\
\rowcolor[HTML]{FFFFFF} 
\multicolumn{1}{c}{\cellcolor[HTML]{FFFFFF}Outliers}   & 24                   & 24                   & 42                   & 42                   \\ \hline
\rowcolor[HTML]{FFFFFF} 
NQC - No Outliers                                      & \textbf{0.330}       & \textbf{0.310}*       & \textbf{0.246}       & \textbf{0.227}       \\
\rowcolor[HTML]{FFFFFF} 
NQC - All                                              & 0.097                & 0.051                & 0.219                & 0.174                \\ \hline
\rowcolor[HTML]{FFFFFF} 
UQC - No Outliers                                       & \textbf{0.350}       & \textbf{0.351}       & \textbf{0.382}       & \textbf{0.340}       \\
\rowcolor[HTML]{FFFFFF} 
UQC - All                                              & 0.206                & 0.151                & 0.333                & 0.283                \\ \hline
\rowcolor[HTML]{FFFFFF} 
WIG - No Outliers                                      & -0.015               & -0.007               & 0.059                & -0.020               \\
\rowcolor[HTML]{FFFFFF} 
WIG - All                                              & 0.077                & 0.041                & 0.002                & -0.103               \\ \hline
\rowcolor[HTML]{FFFFFF} 
QF - No Outliers                                        & \textbf{0.357}       & \textbf{0.369}       & \textbf{0.493}       & \textbf{0.438}       \\
\rowcolor[HTML]{FFFFFF} 
QF - All                                               & 0.283                & 0.281                & 0.469                & 0.403                \\
                                                       & \multicolumn{1}{l}{} & \multicolumn{1}{l}{} & \multicolumn{1}{l}{} & \multicolumn{1}{l}{} \\
                                                       & \multicolumn{1}{l}{} & \multicolumn{1}{l}{} & \multicolumn{1}{l}{} & \multicolumn{1}{l}{} \\
                                                       & \multicolumn{1}{l}{} & \multicolumn{1}{l}{} & \multicolumn{1}{l}{} & \multicolumn{1}{l}{} \\
                                                       & \multicolumn{1}{l}{} & \multicolumn{1}{l}{} & \multicolumn{1}{l}{} & \multicolumn{1}{l}{} \\
                                                       & \multicolumn{1}{l}{} & \multicolumn{1}{l}{} & \multicolumn{1}{l}{} & \multicolumn{1}{l}{}
\end{tabular}
}

\label{Tab:Results}
\end{table}

\section{Conclusion}\label{sec:conclusion}
Studies on QPP generally focus on prediction accuracy~\cite{grivolla2005automatic,zhou2007query,carmel2010estimating,shtok2012predicting,raiber2014query,chifu2018query,faggioli2023query}, but seldom on the difficulty of that prediction. QPP is clearly a difficult task, since current predictors are not very accurate. 
In this study we show that  difficult to predict queries can be detected. We used multivariate outlier detection for that; it has the advantage to consider different QPPs to detect the queries for which the prediction may not be accurate. We also show that removing these automatically detected queries, we have a higher accuracy of the predictor. That means that we know that the predictor is not accurate for some queries that we can identify. This result pave the way to a new research direction: the prediction of the accuracy of the prediction. Some predictor may be accurate for certain queries and not for others. In future work, we will re-examine the results on more benchmark collections and reference retrieval systems. In this study we considered a single feature at a time to make the prediction. In future work, we will consider trained models that combine several features. That will also allow us to consider methods that allow models to abstain from making a prediction when they are not sufficiently confident. 

\bibliographystyle{splncs04}
\bibliography{bib}

\end{document}